\documentclass[10pt]{article}

 \usepackage{amsmath,amsthm,amssymb}
 \usepackage{makeidx,epsfig,lscape}
 \usepackage{color,colortbl}
 \usepackage{fancyhdr}
 \usepackage{xcolor,pict2e}

 \renewcommand{\headrulewidth}{0pt}
 \renewcommand{\footrulewidth}{0.5pt}

 \definecolor{myaqua}{rgb}{0.0,0.5,0.55}
 \definecolor{lightaqua}{rgb}{0.75,0.95,0.95}

 \usepackage[colorlinks = true,
            linkcolor = myaqua,
            urlcolor  = blue,
            citecolor = myaqua]{hyperref}

\usepackage{caption}
\usepackage{floatrow}
 \usepackage{subcaption}

\theoremstyle{definition}
 \newtheorem{defn}{Definition}[section]

\def\lin#1#2{\textcolor[rgb]{0.6,0.6,0.6}{\vspace*{#1mm} \hrule
   height 3 pt \vspace*{#2mm}}}
\def\bt{\begin{tabular}}
\def\et{\end{tabular}}
\def\and{\mbox{ and }}

\def\1{{\bf 1}}

 \def\sectionn#1{\refstepcounter{section}{\color{myaqua}

 \vskip 6mm

 \noindent\Large\bf\thesection. #1}

 \vskip 3mm}
 \def\subsectionn#1{\refstepcounter{subsection}{\color{myaqua}

 \vskip 5mm

 \noindent\large\bf\thesubsection. #1}

 \vskip 2mm}

\begin{document}

 \fancyhead[L]{\hspace*{-13mm}
  \bt{l}{\bf Manuscript}\\
 \et}
 \fancyhead[R]{\includegraphics{pic1.ps}}

 $\mbox{ }$

 \vskip 10mm

{ 

{\noindent{\huge\bf\color{myaqua}
 Investigation of Fractional Compartmental \\ Models with Application to Amiodarone Drug Diffusion in  Pharmacokinetics  }}
\\
%
{\large\bf Reindorf Nartey Borkor$^{1,2}$, Adu Sakyi$^{1,2}$, Peter Amoako-Yirenkyi$^{1,2}$  }
\\[2mm]
{ 
$^1$Department of Mathematics, Kwame Nkrumah University of Science and Technology (KNUST), Ghana\\[1mm]
$^2$Center for Scientific and Technical Computing, National Institute for Mathematical Sciences (NIMS), Ghana\\[1mm]
Email:
\href{mailto:rborkor@nims.edu.gh}{\color{blue}{\underline{\smash{reinbork@knust.edu.gh}}}}
\href{mailto:asakyi@knust.edu.gh}{\color{blue}{\underline{\smash{asakyi@knust.edu.gh}}}},
\href{mailto:amoakoyirenkyi@knust.edu.gh}{\color{blue}{\underline{\smash{amoakoyirenkyi@knust.edu.gh}}}}
\\[3mm]

\lin{5}{7}

 { 
 {\noindent{\large\bf\color{myaqua} Abstract}{\bf \\[2mm]

This paper presents three fractional models formulated from a  classical Pharmacokinetics compartmental system:   commensurable,  non-commensurable,  and implicit non-commensurable models.  Their distinguishing characteristics are further examined comprehensively.  Because analytic solutions for such models are typically challenging to obtain,  we study the application of the Fractional Finite Difference Method (FFDM) to simulate approximate solutions.  The characteristic of the non-commensurable model is shown to be incompatible with the concept of mass balance.   However,  it appeared to outlast fractional calculus theory when simulating anomalous kinetics.  We proved this by fitting the proposed fractional and classical models to an experimental data set (amiodarone) and estimated the parameters using the least-square approach.  The classical model diverged, but the non-commensurable model predicted a fit comparable to the other two fractional models.  The fractional models described anomalous diffusion better than classical theories.  The numerical results showed that the proposed numerical method is equally efficient in solving any complex compartmental models, as they performed well in simulations for the classic example of the model.

 }}

 {\noindent{\large\bf\color{myaqua} Keywords}{\bf \\[3mm]
 Caputo Fractional Derivative; Fractional Finite Difference Methods; Pharmacokinetics; Compartmental analysis
}


\lin{2}{1}

\sectionn{Introduction}
\label{intro}
{ \fontfamily{times}\selectfont
 \noindent 
In the majority of application-oriented fields, compartmental analysis has been a crucial technique. It was initially developed as a result of studies on the absorption and dispersion of radioactive tracers, and it now plays a significant role in a variety of different disciplines, including medicine, bioengineering, environmental science, information science, and social science, to name just a few \cite{pet}. Compartmental analysis appears to have a lengthy history in science according to the literature, but throughout the years, it has grown more well-liked in the field of health science.

Most scientists have conducted extensive research into the usage of compartments in the fields of biology and medicine.
For instance, \cite{Yu} employed a GPU-accelerated compartmental model to analyze medical imaging data, and \cite{Akh} used it to analyze various characteristics of the blood pressure distribution. It has been used to describe and analyze the spread of communicable diseases like measles, coronavirus, ebola, influenza, and tuberculosis in the field of epidemiology (to name a few, see \cite{ivorra,esra,berge,opoku}).
Compartmental models are employed in several specialist domains, such as pharmacology and pharmacokinetics, to forecast the most secure and efficient drug administration method. Drugs are transmitted into and out of these compartments through diffusion (a transport phenomenon), which is a representation of many body parts (such as the stomach, blood, liver, and kidney).
Traditionally, researchers analyze pharmacokinetics data using compartmental and non-compartmental models. Yang et al's analysis of the blood alcohol content of Chinese participants in Hong Kong used both compartmental and generalized linear models \cite{Yang}.

 \renewcommand{\headrulewidth}{0.5pt}
\renewcommand{\footrulewidth}{0pt}

 \pagestyle{fancy}
 \fancyfoot{}
 \fancyhead{} 
 \fancyhf{}
 \fancyfoot[C]{\leavevmode
 \put(0,0){\color{lightaqua}\circle*{34}}
 \put(0,0){\color{myaqua}\circle{34}}
 \put(-2.5,-3){\color{myaqua}\thepage}}

 \renewcommand{\headrule}{\hbox to\headwidth{\color{myaqua}\leaders\hrule height \headrulewidth\hfill}}

 

By means of a set of differential equations, compartmental analysis has traditionally explained how materials are moved between compartments of a system.
Recent developments in compartmental analysis have made it possible to use fractional calculus in the disciplines indicated above as well as others (see \cite{chris,bash,angsn,card}).
In general, modeling systems involving memory (history) and/or non-localized effects can be aided by the use of fractional calculus.
The use of fractional calculus as a modeling technique has increased.
\cite{Par} proposed a dynamical fractional order HIV-1 model in the Caputo sense, establishing the significance of the fractional derivative on dynamic processes, and \cite{Dum} examined the fractional properties of a harmonic oscillator with position-dependent mass as a few examples.
See \cite{bale,amin,sama,bale1,Amin,moha} and the references therein for current advancements in the field of fractional calculus and its applications.
Furthermore, the use of fractional calculus has produced a number of difficult to solve analytically complex mathematical models within the context of differential equations. The generalized Adams-Bashforth-Moulton Method \cite{kara,naik,par}, the fractional finite difference method \cite{swei}, the GL-based method \cite{Zhen}, the Adomian Decomposition Method \cite{ja}, the Chebyshev spectral method \cite{owola}, the collocation method \cite{ra,angel}, and the hybrid Chelyshko functions Method \cite{Moh}, and an Iterative method for fractional optimal control problems \cite{Amin} are some of the numerical techniques that are
employed in finding approximate solutions to these rising complex
 equations in the last decade
For an in-depth analysis of the approximation techniques, see \cite{sam1}.

Although there are other theories, including Fractal Kinetics \cite{marsh,fuite,hickey}, Empirical Power-Laws \cite{wise,panos1}, and Gamma Functions \cite{tucker,weso,weso1}, it is demonstrated in the theory of pharmacokinetics that fractional calculus is the most effective method for explaining the anomalous behavior of some medications.
Dokoumetzidis et al. \cite{dok2} first discussed it in Pharmacokinetics (PK) for a single compartmental model.
Since the method created for fractionalizing a single compartmental system fails in the case of multi-compartmental systems, Dokoumetzidis et al. later devised a theoretically-based method of fractionalizing a multi-compartmental system \cite{dok1}.
This approach poses certain difficulty in interpreting the units of constants if not implemented carefully, and the conservation of mass may be void.
These are the challenges that Dokoumetzidis et al. \cite{dok2} and Angstmann et al. \cite{angs} have discovered and addressed.
Yanli et al. \cite{yan} recently published a paper in which they developed a two compartmental fractional model based on the methodology of Dokoumetzidis et al. and studied two numerical techniques together with parameter estimation.

However, since there is some justification for incorporating historical impacts into the dynamics of compartmental models, researchers frequently adopt either approach.
This study compares the approaches that are applied to a classical model to produce the commensurable, non-commensurable, and implicit non-commensurable fractional models.
Additionally, it is challenging to put analytical solutions for these FDE systems into practice.
Because of this, this study effectively illustrates how to use the Fractional Finite Difference Method (FFDM) to generate a numerical solution for the described systems.
Further, we use the least-squares method to estimate the parameters of the classical and fractional models using an amiodarone drug dataset in order to determine which of them fails to fit the drug's anomalous diffusion behavior.

The rest of this work is structured as follows: In section 2, we give two schematic illustrations of two compartmental systems with a pharmacokinetics application, from which the various models were built.
Additionally, we offer some helpful definitions for fractional calculus, particularly for the caputo derivative. The findings of the numerical simulation are further examined in Section 3 along with the key variations among the models under consideration.
Also, parameter estimation is done using the drug amiodarone.
Finally, we offered conclusions  in section 4.

\sectionn{Methodology}
 { \fontfamily{times}\selectfont
  \noindent
\subsectionn{Single Dose Two-Compartmental Intravenous(IV) Model}\label{I3}
In pharmacokinetics,  a two-compartment model will consist of system of differential equation that represents two physiological essential parts (see a schematic diagram
shown in $figure(\ref{fig1:a})$.):
\begin{itemize}
 \item The first (central) compartment denoted as $A_1$ is marked as  the blood and organs with enough
blood such as kidney or liver.
\item The second (peripheral) compartment also denoted as $A_2$ can be marked as tissue or generally, any body part with less supply of blood.
\end{itemize}

\noindent These compartments are joined to  one another in both directions and hence, a distribution
between the $A_1$ and $A_2$ takes place.

From $figure(\ref{fig1:a})$, a linear model is obtained and are expressed by the following
system of Ordinary Differential Equations (ODE's).

%

 \begin{figure}[!htp] 
  \begin{subfigure}[b]{0.45\linewidth}
    \centering\caption{} 
    \includegraphics[width=\textwidth]{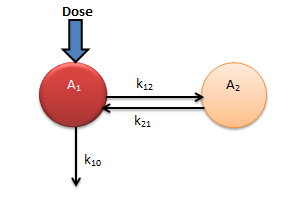} 
    
    \label{fig1:a} 
  \end{subfigure}
  \begin{subfigure}[b]{0.45\linewidth}
    \centering \caption{}
    \includegraphics[width=\textwidth]{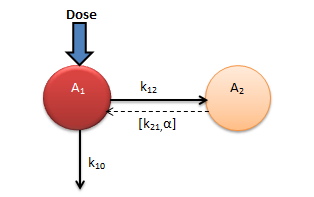} 
    
    \label{fig1:b} 
  \end{subfigure}
 
   \caption{Both diagrams shows general schematic diagram of two-compartmental models with dose administered into the central compartment 
   intravenously. 
 (\ref{fig1:b}) is a fractional 2-compartment PK model with the dashed line representing the fractionalized process from compartment 2 to compartment 1.}
  \label{fig1} 
\end{figure}

 \begin{eqnarray}\nonumber
  \label{eq1}
\frac{dA_1(t)}{dt}&=&-k_{10}A_1(t)-k_{12}A_1(t)+k_{21}A_2(t)\\ 
\frac{dA_2(t)}{dt}&=&k_{12}A_1(t)-k_{21}A_2(t)
  \end{eqnarray}
 
\noindent where $A_1(t)$ and $A_2(t)$ represent the amount of drug in a compartments, the rate
constants ($k_{12},k_{21},k_{10}$) regulate the transfer of mass between compartments and out of a compartment.
 The thick arrow shows initial values of the amount of drug(dose) given
intravenously at time zero into the blood circulation ($ie A_1(0) = dose$ and $A_2(0) = 0$).\\

The idea of Fractional Calculus was assumed by Leibniz, which was written in letter to a colleague in 1695.
In recent years,  models of FDEs have greatly been considered in different research areas, complementing our understanding of the way we observe things
that are previous modeled with the traditional calculus.
The most important property of the fractional models  is their non-local
(memory) property which does not occur in the case of differential operators of integer order.
By this property, we mean a model's next stage depends not only on its current state but also its historical states. (See the following literature for the many definitions and theory of fractional calculus,
\cite{pod},\cite{sam},\cite{rkh}.)
The fractional model will be in the sense of Caputo’s fractional derivative since it
requires an initial condition that involves the state variable.

 \begin{defn}
\textit{The Caputo fractional derivative operator $~^C\!D^{\alpha}$ of order $\alpha$ is defined in the following form:
 }\end{defn}

\begin{equation}
\label{eq2}
~^C_0\!D_x^{\alpha}f(x)= \frac{1}{\Gamma(m-\alpha)}\int_0^x \frac{f^{(m)}(t)}{(x-t)^{\alpha-m+1}}, \qquad \alpha>0,
\end{equation}\\
\textit{where} $m-1<\alpha\leq m, m \in N , x>0.$\\

 Caputo fractional derivative operator is a linear
operation similar to the integer-order derivative\\
\begin{equation}
\label{eq3}
  ~^C_0\!D_x^{\alpha}(\lambda f(x)+\mu g(x))=\lambda ~^C_0\!D^{\alpha}_x f(x)+\mu ~^C_0\!D^{\alpha}_xg(x),
\end{equation}\\
where $\lambda$ and $\mu$ are constants. Note that the Caputo
differential operator collapses to the classical derivative operator if $\alpha \in N$.
The initial conditions for fractional differential equations with the Caputo derivative have the same form as integer-order differential equations, hence this fractional definition is superior for most physical processes. 

\subsectionn{Derivation of Fractional Pharmacokinetics(PK) Models}\label{I4}
\subsubsection{Commensurate Fractional Two-compartmental PK Model}\label{I5}

A general Commensurate fractional compartmental system is of the form
\cite{bon},\cite{vero}:\\

\begin{equation}
 \label{eq4}
 ~^C_0\!D_t^{\alpha}\textbf{y}(t)
=\left(\begin{array}{c} ~^C_0\!D_t^{\alpha}y_1(t) \\ \cdots \\  ~^C_0\!D_t^{\alpha}y_m(t) \end{array}\right)
=\left(\begin{array}{ccc} a_{11} & \cdots & a_{1m} \\ \cdots &\cdots& \cdots \\  a_{m1} &\cdots& a_{mm} \end{array}\right)
\textbf{y}(t)+\textbf{f}(t)=A\textbf{y}(t)+\textbf{f}(t)
 \end{equation}\\
 
\noindent where $\alpha>0$. The initial conditions $\textbf{y}(0) = \textbf{y}_0$
and $\textbf{f}(t)$ is the (vector valued) input function to the system. These systems are defined
as fractionalization of compartments and are termed commensurate. This is because all the derivatives in the
equations are of the same order, $\alpha$. Also, the commensurate system (\ref{eq4}) has a consistency
with units of the rate constants or no violation of mass balance.

Hence, the system of ODEs defined in (\ref{eq1}) is transformed into the following commensurate fractional PK
model of equal order Caputo sense:

 \begin{eqnarray}\nonumber
  \label{eq5}
~^C_0\!D_t^{\alpha}A_1(t)&=&-k_{10}A_1(t)-k_{12}A_1(t)+k_{21}A_2(t)\\ 
~^C_0\!D_t^{\alpha}A_2(t)&=&k_{12}A_1(t)-k_{21}A_2(t)
  \end{eqnarray}
where $k_{12},k_{21}$ and $k_{10}$ are the rate constants with unit (time$^{-\alpha}$)
  \subsubsection{Non-Commensurate Fractional Two-compartmental PK model}
Also, a general Commensurate fractional compartmental system is of the form \cite{bon},\cite{vero}:\\

\begin{equation}
 \label{eq6}
\left(\begin{array}{c} ~^C_0\!D_t^{\alpha_1}y_1(t) \\ \cdots \\  ~^C_0\!D_t^{\alpha_m}y_m(t) \end{array}\right)
=\left(\begin{array}{ccc} a_{11} & \cdots & a_{1m} \\ \cdots &\cdots& \cdots \\  a_{m1} &\cdots& a_{mm} \end{array}\right)
\textbf{y}(t)+\textbf{f}(t)=A\textbf{y}(t)+\textbf{f}(t)
 \end{equation}

\noindent where $\alpha_1,\cdots,\alpha_m$ are different positive real values representing the non-integer order of each compartment.
Similarly, the system of ODEs defined in (\ref{eq1}) is transformed into Non-commensurate
fractional PK model same way as (\ref{eq5}) hence, also defined as fractionalization of compartments but
with different order.\\
\begin{eqnarray}\nonumber
  \label{eq7}
~^C_0\!D_t^{\alpha}A_1(t)&=&-k_{10}A_1(t)-k_{12}A_1(t)+k_{21}A_2(t)\\ 
~^C_0\!D_t^{\beta}A_2(t)&=&k_{12}A_1(t)-k_{21}A_2(t)
  \end{eqnarray}
  
\noindent There is a major problem to Non-commensurate systems. They have features which contradict mass
balance. (ie, the units of the rate are inconsistencies.)  Thus, a mass flux leaving one compartment with 
defined rate of non-integer order, appears as a mass flux entering into another compartment, as a rate
of a different non-integer order, and as such violates mass balance\cite{dok3}.

\subsubsection{Implicit Non-Commensurate Fractional Two-compartmental PK model}
We consider the form of fractionalizing a system of ODE 
which takes care of the problem encountered with Non-commensurate systems\cite{dok1}. Unlike the previous
fractionalized forms where each compartment or equation is fractionalized with equal (or unequal)
order, this approach considers separate fractionalization of each transport process of the system.
In this case, processes of distinct fractional orders co-exist. And this is because, they are of consistent orders when the
corresponding terms show up in different equations. In addition, this approach does not encounter any
problem of mass  imbalance.

Now, for a fractionalized transfer processes in $figure(\ref{fig1:b})$, the generalized fractional form of the
two-compartmental systems in Caputo's derivative sense is given as [see \cite{dok1}
for more details on how the fractionalization on any system is done]:

\begin{eqnarray}\nonumber
  \label{eq8}
\frac{dA_1(t)}{dt}&=&-k_{10}~^C_0\!D_t^{1-\alpha_{10}}A_1(t)-k_{12}~^C_0\!D_t^{1-\alpha_{12}}A_1(t)+k_{21}~^C_0\!D_t^{1-\alpha_{21}}A_2(t)\\ 
\frac{dA_2(t)}{dt}&=&k_{12}~^C_0\!D_t^{1-\alpha_{12}}A_1(t)-k_{21}~^C_0\!D_t^{1-\alpha_{21}}A_2(t)
  \end{eqnarray}

where $k_{10},k_{12}$ and $k_{21}$ are the rate constants with units as time$^{\alpha_{10}}$, time$^{\alpha_{12}}$ and time$^{\alpha_{21}}$ respectively.
\noindent With a schematic diagram shown in $figure(\ref{fig1:b})$,
a well perfused tissues could represent $A_1$ while $A_2$ as deeper tissues.
Also, with
three transfer rates denoted as $k_{10},k_{12}$ and $k_{21}$, we assume classical kinetics(order 1) for $k_{10}$ and
$k_{12}$ (that is $\alpha_{10} = \alpha_{12} = 1$ from (\ref{eq8}) but fractional kinetics(order $\alpha$) for $k_{21}$ signifying tissue
trapping(dashed arrow). The thick arrow shows the amount of drug(dose) given intravenously at
time zero into the blood circulation (compartment $A_1$).

The system (\ref{eq8}) is mathematically deduced as follows:

\begin{eqnarray}\nonumber
  \label{eq9}
\frac{dA_1(t)}{dt}&=&-(k_{10}+k_{12})A_1(t)+k_{21}~^C_0\!D_t^{1-\alpha}A_2(t)\\ 
\frac{dA_2(t)}{dt}&=&k_{12}A_1(t)-k_{21}~^C_0\!D_t^{1-\alpha}A_2(t)
  \end{eqnarray}
\noindent where $\alpha < 1$.  Furthermore, $A_1(0) = dose$ and $A_2(0) = 0$ are the initial conditions that account for a bolus
dose injection in $A_1$ and no initial amount in $A_2$, respectively. We refer (\ref{eq8}) and (\ref{eq9}) as an
implicit form of a system of fractional Pharmacokinetics(PK) models. In the subsequent sections, we
derive numerical methods in finding solutions to the above pharmacokinetics models since analytical
solutions cannot easily be implemented especially (\ref{eq7}) and (\ref{eq9}).\\

\subsectionn{Numerical Approximation and Simulation}

Since  exact analytic solutions of some fractional order differential equations are difficult to obtain, a
 numerical approximation must be used. In solving fractional order equations, several numerical schemes
have been proposed. (See section \ref{intro} for the list of methods.) For numerical solutions of the
fractional systems, we consider Fractional Finite Difference Method (FFDM) in this work. We demonstrate the
FFDM on only model (\ref{eq9}) which in similar terms can be used for the other models (\ref{eq5},\ref{eq7}).\\

\subsubsection{Approximation of Caputo Fractional Derivative}

Here, a fractional derivative $D^{\alpha}u(t)$ in a discrete approximation form is presented. For a given
positive integer $M$ (defining the number of grids), finite difference algorithm 
in the time interval $[0,T_f]$
is defined by $k =\frac{T_f}{M}$ with grid points labeled as $t_n = nk, n = 0, 1, 2, ..., M$.

The discrete approximation of $D^{\alpha}u(t)$ is derived by a simple
quadrature formula. (See \cite{swei} for more details.):

\begin{eqnarray}\nonumber
  \frac{d^{\alpha}u(t_n)}{dt^{\alpha}}&=&\frac{1}{\Gamma(1-\alpha)}\int_0^{t_n} (t_n-s)^{-\alpha}\frac{d}{ds}u(s)ds\\ \nonumber
                                 &=&\frac{1}{\Gamma(1-\alpha)}\sum_{j=1}^n\int_{(j-1)k}^{jk}[\frac{u_j-u_{j-1}}{k}+o(k)](nk-s)^{-\alpha}ds\\ \nonumber
                                 &=&\frac{1}{\Gamma(1-\alpha)(1-\alpha)}\sum_{j=1}^n[\frac{u_j-u_{j-1}}{k}+o(k)]
                                 [(n-j+1)^{1-\alpha}-(n-j)^{1-\alpha}][k^{1-\alpha}]\\ \nonumber
                                 &=&\frac{1}{\Gamma(1-\alpha)(1-\alpha)k^{\alpha}}\sum_{j=1}^n (u_j-u_{j-1})[(n-j+1)^{1-\alpha}-(n-j)^{1-\alpha}]\\ \nonumber
                                 & & +\frac{1}{\Gamma(1-\alpha)(1-\alpha)}\sum_{j=1}^n [(n-j+1)^{1-\alpha}-(n-j)^{1-\alpha}]o(k^{2-\alpha}). \nonumber
 \end{eqnarray}
Now, we set  and shift indices to give: 
\begin{eqnarray} 
\label{eq10}
 \sigma_{\alpha,k}=\frac{1}{\Gamma(1-\alpha)(1-\alpha)k^{\alpha}}\\ 
 \label{eq11}
 \omega_j^{(\alpha)}=j^{1-\alpha}-(j-1)^{1-\alpha}
\end{eqnarray}
and
  \begin{eqnarray} \nonumber
   \frac{d^{\alpha}u(t_n)}{dt^{\alpha}}&=& \sigma_{\alpha,k}\sum_{j=1}^n \omega_j^{(\alpha)}(u_{n-j+1}-u_{n-j})+\frac{1}{\Gamma(1-\alpha)(1-\alpha)}n^{1-\alpha}o(k^{2-\alpha})\\ \nonumber
                                        &=& \sigma_{\alpha,k}\sum_{j=1}^n \omega_j^{(\alpha)}(u_{n-j+1}-u_{n-j}) +o(k).\nonumber
   \end{eqnarray}
Here
\begin{eqnarray} \nonumber
 \frac{d^{\alpha}u(t_n)}{dt^{\alpha}}=D^{\alpha}u_n+o(k) \nonumber
\end{eqnarray}
and the first-order approximation for  computing the  Caputo's fractional derivative is expressed as
\begin{equation}
\label{eq12}
 D^{\alpha}u_n\cong \sigma_{\alpha,k}\sum_{j=1}^n \omega_j^{(\alpha)}(u_{n-j+1}-u_{n-j}) , \qquad n=1,2,...,M.
\end{equation}

\subsectionn{Discretizing System of Fractional PK Model using Fractional Finite
Difference Method(FFDM)}
The discretized formula of the FFDM  (\ref{eq12}) now approximate  
time $\alpha$-order fractional derivative to solve the fractional PK model (\ref{eq9}) numerically. The discretized
form is presented as follows;

\begin{eqnarray}\nonumber
\label{eq13}
 \frac{A^1_n-A^1_{n-1}}{k}&=&-(k_{10}+k_{12}) A^1_n +k_{21}\delta_{1-\alpha,k}\sum_{j=1}^{n} w^{(1-\alpha)}_j(A^2_{n-j+1}-A^2_{n-j})\\ 
 \frac{A^2_n-A^2_{n-1}}{k}&=&k_{12}A^1_n -k_{21}\delta_{1-\alpha,k}\sum_{j=1}^{n} w^{(1-\alpha)}_j(A^2_{n-j+1}-A^2_{n-j})
\end{eqnarray}

\noindent where for instance $A^1_n=A_1(t_n)$ and $A^2_{n-j+1}=A_2(t_{n-j+1})$ and $k$ is the interval between grid points.
From expression (\ref{eq10}) and (\ref{eq11}), $\delta_{1-\alpha,k}$ and $w_j^{(1-\alpha)}$in (\ref{eq13})  are represented as:

\begin{eqnarray} 
\label{eq14}
 \delta_{1-\alpha,k}&=&\frac{1}{\alpha k^{1-\alpha}\Gamma(\alpha)}\\ 
 \label{eq15}
 w_j^{(1-\alpha)}&=&j^{\alpha}-(j-1)^{\alpha}
\end{eqnarray}

\noindent Hence,
\begin{eqnarray}\nonumber
\label{eq16}
 A^1_n-A^1_{n-1}&=&-k(k_{10}+k_{12}) A^1_n +k k_{21}\delta_{1-\alpha,k}\sum_{j=1}^{n} w^{(1-\alpha)}_j(A^2_{n-j+1}-A^2_{n-j})\\ 
 A^2_n-A^2_{n-1}&=&kk_{12}A^1_n -kk_{21}\delta_{1-\alpha,k}\sum_{j=1}^{n} w^{(1-\alpha)}_j(A^2_{n-j+1}-A^2_{n-j})
\end{eqnarray}
\noindent Now a couple of iterates are taken to generalize the process.\\

\noindent for $n=1$
\begin{eqnarray}\nonumber
\label{eq17}
 A^1_1-A^1_{0}&=&-k(k_{10}+k_{12}) A^1_1 +k k_{21}\delta_{1-\alpha,k} w^{(1-\alpha)}_1(A^2_{1}-A^2_{0})\\ 
 A^2_1-A^2_{0}&=&kk_{12} A^1_1 -k k_{21}\delta_{1-\alpha,k} w^{(1-\alpha)}_1(A^2_{1}-A^2_{0})
\end{eqnarray}

\noindent Grouping known values and unknown terms, we obtain the following,
\begin{eqnarray}\nonumber
\label{eq18}
(1+k(k_{10}+k_{12}))A_1^1-(k k_{21}\delta_{1-\alpha,k} w^{(1-\alpha)}_1)A_1^2=A_0^1-(k k_{21}\delta_{1-\alpha,k} w^{(1-\alpha)}_1)A_0^2\\ 
-kk_{12}A_1^1+(1+k k_{21}\delta_{1-\alpha,k} w^{(1-\alpha)}_1)A_1^2=A_0^1+(k k_{21}\delta_{1-\alpha,k} w^{(1-\alpha)}_1)A_0^2
 \end{eqnarray}
\noindent Putting (\ref{eq18}) in a matrix form\\

$\left(\begin{array}{cc} {(1+k(k_{10}+k_{12}))} & {-(k k_{21}\delta_{1-\alpha,k} w^{(1-\alpha)}_1)}  \\ {-kk_{12}} & {(1+k k_{21}\delta_{1-\alpha,k} w^{(1-\alpha)}_1)} \end{array}\right)$
$\left(\begin{array}{c} {A_1^1}  \\ {A_1^2}  \end{array}\right)$
=$\left(\begin{array}{c} {A_0^1-(k k_{21}\delta_{1-\alpha,k} w^{(1-\alpha)}_1)A_0^2}  \\ {A_0^1+(k k_{21}\delta_{1-\alpha,k} w^{(1-\alpha)}_1)A_0^2}  \end{array}\right)$  \\

\noindent for $n=2$
\begin{eqnarray}\nonumber
\label{eq19}
 (1+k(k_{10}+k_{12}))A_2^1-(kk_{21}\delta_{1-\alpha,k} w^{(1-\alpha)}_1)A_2^2&=&A_1^1+(kk_{21}\delta_{1-\alpha,k}w^{(1-\alpha)}_2)A_1^2  \\ \nonumber
                                                                         &-&kk_{21}\delta_{1-\alpha,k} (w^{(1-\alpha)}_1 A_1^2 +w^{(1-\alpha)}_2 A_0^2)\\ \nonumber                               
-kk_{12}A_2^1+(1+k k_{21}\delta_{1-\alpha,k} w^{(1-\alpha)}_1)A_2^2&=&A_1^1-(kk_{21}\delta_{1-\alpha,k}w^{(1-\alpha)}_2)A_1^2\\
                                                                       &+& kk_{21}\delta_{1-\alpha,k} (w^{(1-\alpha)}_1 A_1^2 +w^{(1-\alpha)}_2 A_0^2)
 \end{eqnarray}
 
 \noindent Also putting (\ref{eq19}) in a matrix form

   \begin{multline} 
   \label{eq20}
   \left(\begin{array}{cc} {(1+k(k_{10}+k_{12}))} & {-(k k_{21}\delta_{1-\alpha,k} w^{(1-\alpha)}_1)}  \\ -kk_{12} & {(1+k k_{21}\delta_{1-\alpha,k} w^{(1-\alpha)}_1)} \end{array}\right)\left(\begin{array}{c} {A_2^1}  \\ {A_2^2}  \end{array}\right)\\
   =\left(\begin{array}{c} {A_1^1+(k k_{21}\delta_{1-\alpha,k} w^{(1-\alpha)}_2)A_1^2-k k_{21}\delta_{1-\alpha,k}\sum_{j=1}^2 w^{(1-\alpha)}_jA^2_{2-j}}  \\ {A_1^1-(k k_{21}\delta_{1-\alpha,k} w^{(1-\alpha)}_2)A_1^2+k k_{21}\delta_{1-\alpha,k}\sum_{j=1}^2 w^{(1-\alpha)}_jA^2_{2-j}}  \end{array}\right)
   \end{multline}

\noindent Generalizing the iterate follows as: 

\begin{multline}\label{eq21}
\left(\begin{array}{cc} {(1+k(k_{10}+k_{12}))} & {-(k k_{21}\delta_{1-\alpha,k} w^{(1-\alpha)}_1)}  \\ {-kk_{12}} & {(1+k k_{21}\delta_{1-\alpha,k} w^{(1-\alpha)}_1)} \end{array}\right)\left(\begin{array}{c} {A_n^1}  \\ {A_n^2}  \end{array}\right)\\
=\left(\begin{array}{c} {A_{n-1}^1+k k_{21}\delta_{1-\alpha,k} \sum_{j=1}^{n-1}w^{(1-\alpha)}_{j+1}A_{n-j}^2-k k_{21}\delta_{1-\alpha,k}\sum_{j=1}^n w^{(1-\alpha)}_jA^2_{n-j}}  \\ {A_{n-1}^1-k k_{21}\delta_{1-\alpha,k} \sum_{j=1}^{n-1}w^{(1-\alpha)}_{j+1}A_{n-j}^2+k k_{21}\delta_{1-\alpha,k}\sum_{j=1}^n w^{(1-\alpha)}_jA^2_{n-j}}  \end{array}\right)
\end{multline}
\noindent Equation (\ref{eq21}) is an iterate generalization  of the 
numerical solutions to the system of fractional PK model (\ref{eq9}), which can
be obtained when grid interval is defined.

}

\sectionn{Results and Discussion}
\label{sec:MLE}


{ \fontfamily{times}\selectfont
 \noindent


The two-compartmental PK models, which are all systems of fractional differential equations, were subjected to a numerical tool (FFDM) in the preceding subsection. With $(\alpha = 1)$ for classical instances and $(\alpha = 0.5, \beta = 0.7)$ for fractional cases, solutions are given using parameter constants specified as $k_{10}= 1, k_{12}= 0.8,$ and $k_{21}= 0.7$. As a result, for model (\ref{eq9}), it is noted that rate constant $k_{21}$ has a unit of $hours^{-0.5}$, which is distinct from the other rate constants ($k_{10}, k_{12}$), which have units of $hours^{-1}$.
The graphs display the time profiles of a certain drug's concentration (or amount) when it is injected intravenously into a human being. As previously stated, $A_1$ and $A_2$ denoted, respectively, the amount of medication in the blood and deeper tissues.
In these three different fractional model types, a drug's behavior or amount is observed over the course of a day ($(24hrs)$). Two profiles are shown on each model, one of which corresponds to a system compartment. The observed profiles $figure(\ref{fig7:a},\ref{fig7:c},\ref{fig7:e})$ and $figure(\ref{fig7:b},\ref{fig7:d},\ref{fig7:f})$  for compartments $A_1$ and $A_2$ of the various models, respectively, depict the drug's behavior in each compartment. Each profile has an overlap between the numerical solution from FFDM and the analytical solution for the classical situation $(\alpha = 1)$. As long as a solution to a differential equation exists, the FFDM is sufficient to provide a satisfactory approximation.


\begin{figure}[!htp] 
  \begin{subfigure}[b]{0.5\linewidth}
    \centering\caption{} 
    \includegraphics[width=1\linewidth]{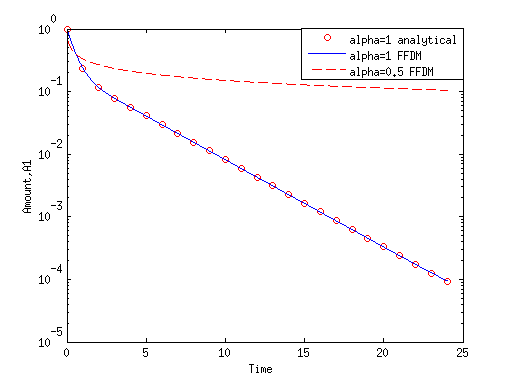} 
    
    \label{fig7:a} 
  \end{subfigure}
  \begin{subfigure}[b]{0.5\linewidth}
    \centering \caption{}
    \includegraphics[width=1\linewidth]{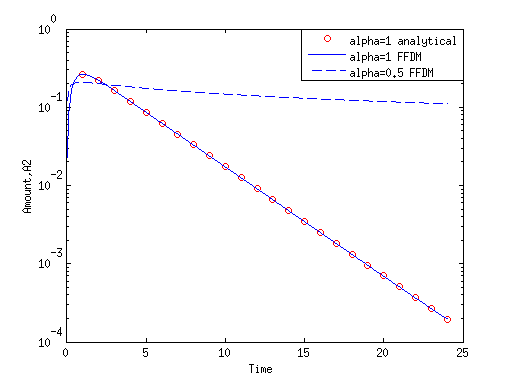} 
    
    \label{fig7:b} 
  \end{subfigure}  
  \begin{subfigure}[b]{0.5\linewidth}
    \centering \caption{} 
    \includegraphics[width=1\linewidth]{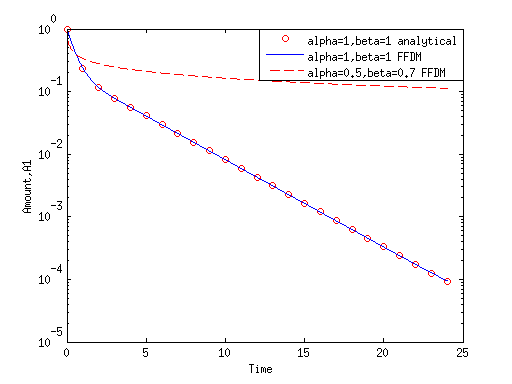} 
   
    \label{fig7:c} 
  \end{subfigure}
  \begin{subfigure}[b]{0.5\linewidth}
    \centering\caption{}
    \includegraphics[width=1\linewidth]{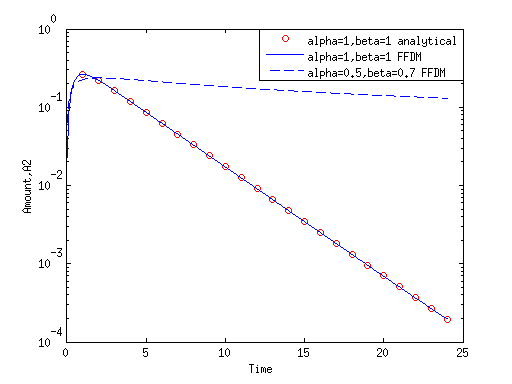} 
     
    \label{fig7:d} 
  \end{subfigure} 
  \begin{subfigure}[b]{0.5\linewidth}
    \centering \caption{} 
    \includegraphics[width=1\linewidth]{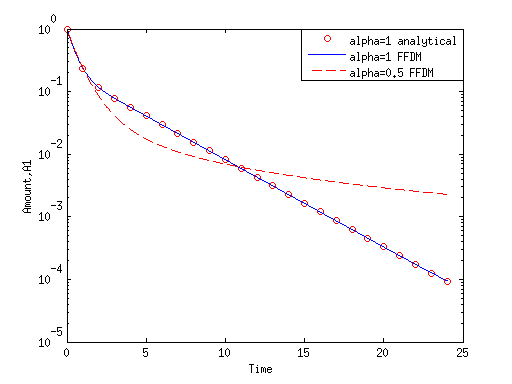} 
   
    \label{fig7:e} 
    \vspace{4ex}
  \end{subfigure}
  \begin{subfigure}[b]{0.5\linewidth}
    \centering \caption{} 
    \includegraphics[width=1\linewidth]{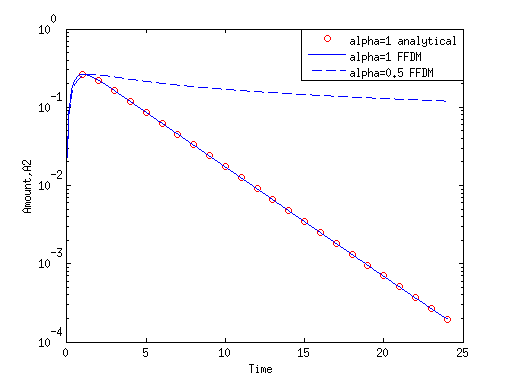} 

    \label{fig7:f} 
    \vspace{4ex}
  \end{subfigure}

  \caption{Time profile of amount of drug in compartment $1 \& 2$ of the three 
   fractional models. Subfigures: (\ref{fig7:a},\ref{fig7:b}),(\ref{fig7:c},\ref{fig7:d}) and (\ref{fig7:e},\ref{fig7:f})
   represent the profile of models $(\ref{eq5})$,$(\ref{eq7})$ and $(\ref{eq9})$ respectively. That is,
   each row represents the behavior of drug in both compartments  under each models respectively.  }
  \label{fig8} 
\end{figure}

\begin{figure}[!htp] 
  \begin{subfigure}[b]{0.5\linewidth}
    \centering\caption{} 
    \includegraphics[width=1\linewidth]{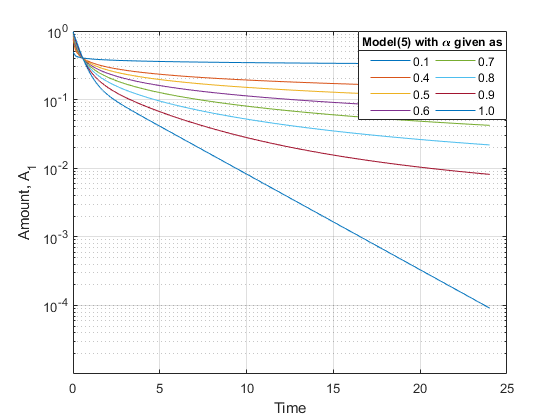} 
    
    \label{fig77:a} 
  \end{subfigure}
  \begin{subfigure}[b]{0.5\linewidth}
    \centering \caption{}
    \includegraphics[width=1\linewidth]{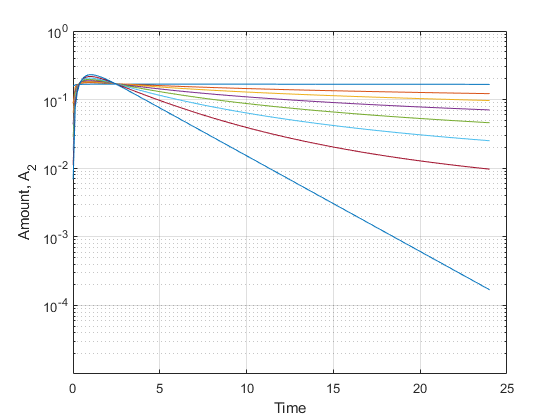} 
    
    \label{fig77:b} 
  \end{subfigure}  
  \begin{subfigure}[b]{0.5\linewidth}
    \centering \caption{} 
    \includegraphics[width=1\linewidth]{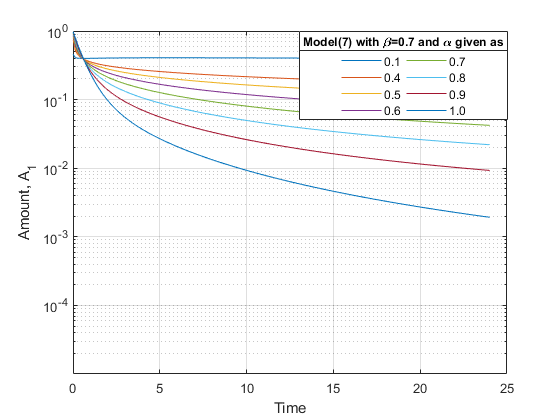} 
   
    \label{fig77:c} 
  \end{subfigure}
  \begin{subfigure}[b]{0.5\linewidth}
    \centering\caption{}
    \includegraphics[width=1\linewidth]{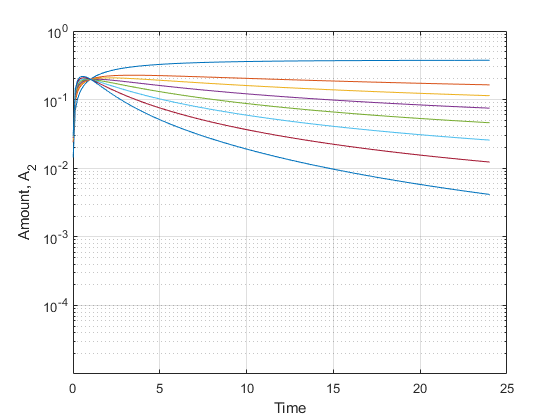} 
     
    \label{fig77:d} 
  \end{subfigure} 
  \begin{subfigure}[b]{0.5\linewidth}
    \centering \caption{} 
    \includegraphics[width=1\linewidth]{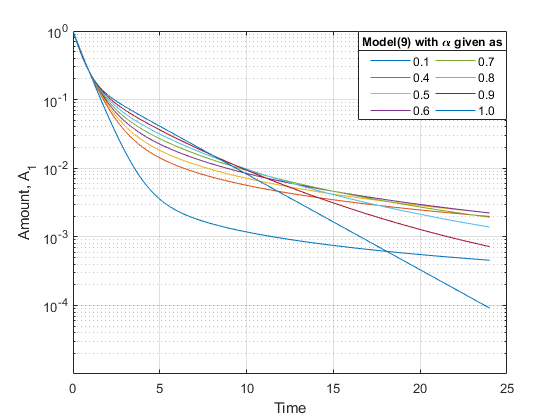} 
   
    \label{fig77:e} 
    \vspace{4ex}
  \end{subfigure}
  \begin{subfigure}[b]{0.5\linewidth}
    \centering \caption{} 
    \includegraphics[width=1\linewidth]{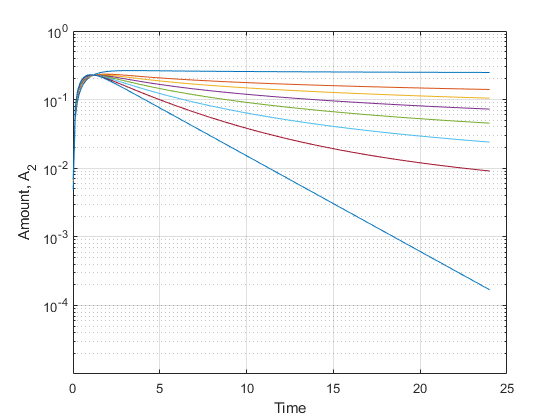} 

    \label{fig77:f} 
    \vspace{4ex}
  \end{subfigure}

  \caption{Time profile of amount of drug in compartment $1 \& 2$ of the three 
   fractional models at various $\alpha$. Subfigures: (\ref{fig77:a},\ref{fig77:b}),(\ref{fig77:c},\ref{fig77:d}) and (\ref{fig77:e},\ref{fig77:f})
   represent the profile of models $(\ref{eq5})$,$(\ref{eq7})$ and $(\ref{eq9})$ respectively.  }
  \label{fig88} 
\end{figure}


In contrast to the classic case, the resulting fractional case in the various profiles typically shows slower, non-exponential dynamics. The fractional profile appeared to have a quicker phase at first but then slowed down. The fractional case in $figure(\ref{fig7:a},\ref{fig7:c})$  exhibits a faster behavior at a very short time (say $1hour$) as opposed to the fractional case in $figure(\ref{fig7:e})$, which also exhibits a faster behavior but takes a longer time (say $10hours$ of the running time) before exhibiting a slower phase. This is another example of the anomalous behavior among the fractional models.
Though all three models employ the same parameter values, it should be noted that some of the parameters have different units.
For instance, because the parameter $k_{12}$ in model $(\ref{eq5})$ has different units ($hours^{-1}$ vs. $hours^{-0.5}$), it cannot be compared across the fractional and classical situations. Additionally, each model equation for each scenario uses the same units for $k_{12}$.
The parameter $k_{12}$, on the other hand, has different units in the two equations in $\ref{eq7})$. For instance, the first and second equations call for the parameter $(k_{12})$ to have a value of $hours^{-0.5}$ and $hours^{-0.7}$, respectively.
However, there is a note we made regarding the figures, particularly $figure(\ref{fig7:a},\ref{fig7:b},\ref{fig7:c},\ref{fig7:d})$.
Despite the model's unit inconsistencies, we found that the non-commensurable system exhibits a comparable tendency to that of the commensurable system.We can also see that compartment 1 behaves similarly for commensurable and non-commensurable systems in $figure(\ref{fig88})$ as $\alpha$ increases compared to the implicit non-commensurate model.
But among all models, the amount of medication in compartment 2 decreases more quickly.

We propose that most medications are distributed over a substantially longer period of time, deviating from the classical case of diffusion to a condition known as anomalous diffusion, which is best represented by power-laws. Instead of the rate constant value, the slower kinetics result from the power-characteristics law's in the terminal phase of the fractional case. We physically connect the slower kinetics to deeper bone and tissue where bone-seeking elements like plutonium, strontium, and calcium exhibit this type of abnormal diffusion behavior. One medication with unusual, non-exponential (power-laws) kinetics is the antiarrhythmic medication \textit{amiodarone}.
To show that fractional differential equation worth for further research in the field pharmacokinetics, 
we consider fitting the Classical PK model  $(\ref{eq1})$  and the Fractional PK models $(\ref{eq5}, \ref{eq7}, \ref{eq9})$ to an amiodarone dataset taken from \cite{weiss} to estimate the parameters of the model $(V_1, k_{10} , k_{12} , k_{21} , \alpha$ and $\beta)$ since in practice, data are primarily available for the central compartment of the system. The outcome of the numerical simulation, $A_1(t)$ is connected to the equation below to determine the parameters.
\begin{equation}
 \label{eq111}
 C(t)=A_1(t)/V_1
\end{equation}
\noindent where $C$ is the concentration of the drug in the blood and $V_1$ is the volume of distribution.

  \begin{figure}[!htp]
 \centering
 \includegraphics[width=360pt]{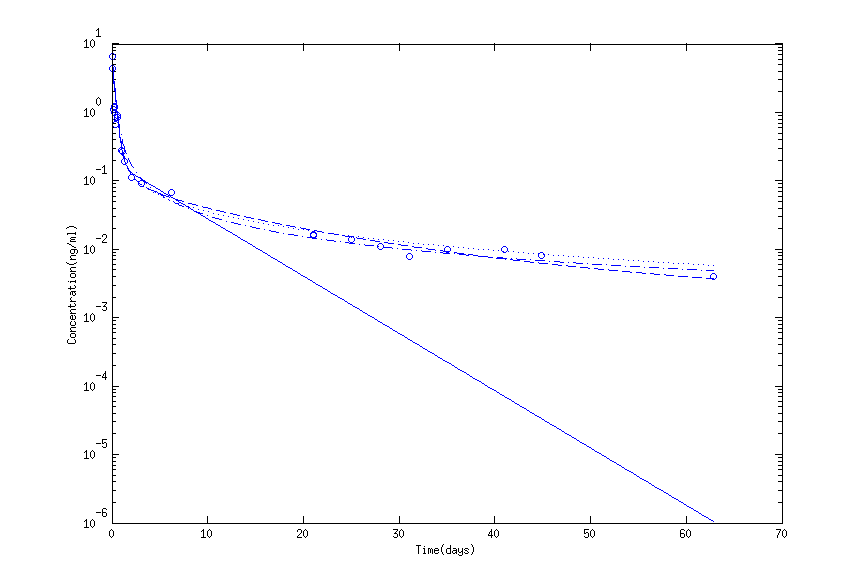}
 \caption{ Experimental data (circles) with superimposed the fit of linear two compartmental system:
ordinary system, $Eq.(\ref{eq1})$ in the text (solid line), commensurate FDE, $Eq.(\ref{eq5})$  (dashed line), non-commensurate FDE, $Eq.(\ref{eq7})$  (dash-dotted line) and implicit non-commensurate FDE, $Eq.(\ref{eq9})$  (dotted
line). \label{fig9}}
\end{figure}

\noindent
$Table(\ref{t1})$ shows the parameter estimates of the various models with their corresponding standard errors. Also, $figure(\ref{fig9})$ displays a semi-log plots of the experimental data with a best fit
of the classical and fractional PK models $(\ref{eq1},\ref{eq5}, \ref{eq7}, \ref{eq9})$ considered under this work. We take a
good look at the unit of these parameters of each model. The classical model $(\ref{eq1})$ has the unit of
all parameters to be $(day^{-1})$ and fractional model $(\ref{eq5})$ with the fractional order as $\alpha$ has unit $(day^{-\alpha})$
for all parameters. Also, fractional model $(\ref{eq9})$ has unit $(day^{-1})$ for parameters $(k_{10} , k_{12})$ and a different
unit $(day^{-\alpha})$ for parameter $(k_{21})$ which poses no deficiency in the model. Moreover, fractional model $(\ref{eq7})$
has unit $(day^{-\alpha})$ for $(k_{10})$ but the parameters $(k_{12} , k_{23})$ have unit $(day^{-\alpha})$ or $(day^{-\beta})$ depending
where they appear in the model (equation) hence inconsistency of units show up.
Nonetheless, the non-commensurate model $(\ref{eq7})$ together with the
other two fractional models $(\ref{eq5}, \ref{eq9})$ give an adequate fit to the experimental data from visual inspection
since amiodarone data are known to have an anomalous behavior (kinetics). 
This is a further observation to the non-commensurable systems that it will equally fit to an experimental data as equal as the other models.
On the other hand, the
classical PK model initially behaves well but later deviates from the anomalous behavior of the drug
since the classical model are always exponential sum. The difference between both kinetics
is that power-law always has slower processes as compared to the exponential no matter how massive the half-life is. The advantage 
of defining the data with Power-law kinetics, is its significant clinical inferences, including infinite
AUC and accumulation without reaching a steady state \cite{dok2}.

 \begin{table}[!htp]
 \centering
  \begin{tabular}{c|c|ccc}
  \hline
\multicolumn{5}{c}{Parameter Estimates} \\
\hline
 {} & {Classical model} & {} & {Fractional model} & {}  \\
     \hline
     {Parameter} & {$model(\ref{eq1})$} & {$model(\ref{eq5})$} & {$model(\ref{eq7})$} & {$model(\ref{eq9})$}  \\
     \hline
   $ k_{10}$  & $3.7598(0.1053) $ &  $ 2.8964(0.4215)$ &   $3.1564(0.6651) $ &   $1.7150(1.0830) $ \\
  $k_{12}$   & $1.2705(0.3442) $ &    $1.2294(0.4212) $ &   $2.1502(0.2632) $ &   $ 2.9483(0.5254)$  \\
  $k_{21}$   & $ 0.2996(0.1381)$ &   $0.1328(0.0660) $ &   $1.2412(0.1829) $ &   $0.3412(0.1310) $  \\
 $\alpha$ & $- $ &   $0.9437(0.0431) $ &  $0.9192(0.2331) $ &  $ 0.5506(0.0268)$  \\
 $\beta$  & $- $ &  $- $ &  $0.4444(0.6390) $ &  $ -$  \\
 $V_1$  &   $ 4.4787(3.4320)$ &  $13.0501(2.2064) $ &  $ 6.0274(2.9343)$ &  $ 9.0547(1.3430)$    \\           
   \hline
  \end{tabular}
  \caption{Parameter estimates (and standard errors) of the PK models fitted to the amiodarone data.\label{t1}}
 \end{table}

 }

\sectionn{Conclusion}
\label{sec:K-M}

{
This work considered modeling  and investigation of  the application of fractional calculus to a classical
compartmental PK model specifically, two compartments in three different ways: (1) A Commensurate
fractional two-compartmental PK model where classical derivatives are simply changed to fractional
of equal real-value order. (2) A Non-Commensurate fractional two-compartmental PK model where
classical derivatives are also changed to fractional of unequal real-value order. (3) A
form of fractionalization where unlike that of the previous two, fractionalization is to the processes not
to the compartment. We termed this form as implicit non-commensurate fractional two-compartmental
PK model. 
We have discovered that fractional PK models often exhibit slower power-law behavior than the classical instance, which is always exponential.
Furthermore, we discovered that compartment fractionalized models, such as the Commensurate and Non-Commensurate PK models, are significantly slower than the implicit non-commensurate PK model, which is owing to the fractionalization of the processes. 
%
Although the non-commensurate PK model does not theoretically give a unit consistency, it should at the very least lead to some anomalies in the results presented in this paper.
However, this particular model results illustrated the required qualities of a fractional model as compared to the others, and hence it represents a state of a system. 
Furthermore, by visual inspection and the output presented in the table, it is seen that non-commensurate model together with the other fractional models fitted well with an amiodarone data
unlike the classical model which deviated. Generally, they can be used to investigate any other typical dataset that shows a complex fractional
kinetics. Also, we demonstrated that the numerical method proposed  work efficiently well for any  other
complex compartmental system. Since 
numerical simulation for the classical case of the system prove to be consistent with the analytical results.

}

 {\color{myaqua}

 \vskip 6mm

 \noindent\Large\bf Acknowledgments}

 \vskip 3mm

{ \fontfamily{times}\selectfont
 \noindent
 Special thanks to National Institute for Mathematical Science (NIMS)
 for its support and contributions. This support is greatly appreciated.\\

 {\color{myaqua}

}}}

\end{document}